\documentclass[a4paper, 11pt,english]{article}

\usepackage[T1]{fontenc}
\usepackage{blindtext}
\usepackage{amsmath}
\usepackage{amstext}
\usepackage{wasysym}
\usepackage{relsize}
\usepackage{esint}
\usepackage{babel}
\usepackage{mathtools}
\usepackage{relsize}
\usepackage{lscape}
\usepackage{cancel}
\usepackage{bm}
\usepackage{bbm}
\usepackage{multirow} 
\usepackage{amsfonts}
\usepackage{amssymb}
\usepackage{graphicx}
\usepackage[usenames,dvipsnames]{color}
\usepackage{hyperref}
\usepackage{epstopdf,epsfig}
\usepackage{verbatim}
\usepackage[utf8]{inputenc}
\usepackage[left=1in,right=1in,top=1in,bottom=1in]{geometry}             % For good copy.
\usepackage{booktabs} % for much better looking tables
\usepackage{array} % for better arrays (eg matrices) in maths
\usepackage{paralist} % very flexible & customisable lists (eg. enumerate/itemize, etc.)
\usepackage{verbatim} % adds environment for commenting out blocks of text & for better verbatim
\usepackage{subfig} % make it possible to include more than one captioned figure/table in a single float
\hypersetup{
    colorlinks=true,         % false: boxed links; true: colored links, false is default
    linkcolor=blue,          % color of internal links, red is default
    citecolor=red,        % color of links to liography, 'green' is default
    urlcolor=Violet             % color of external links, cyan is default
}

\begin{document}

\title{Finite Cutoff AdS$_{5}$ Holography and the Generalized Gradient Flow}
\author{{\bf Vasudev Shyam}\footnote{\href{mailto:vshyam@pitp.ca}{vshyam@pitp.ca}} \\\it Perimeter Institute for Theoretical Physics\\ \it 31 N. Caroline St. Waterloo, ON, N2L 2Y5, Canada }
\date{\today}                                           % Activate to display a given date or no date

\maketitle
\begin{abstract}
Recently proposed double trace deformations of large $N$ holographic CFTs in four dimensions define a one parameter family of quantum field theories, which are interpreted in the bulk dual as living on successive finite radius hypersurfaces. The transformation of variables that turns the equation defining the deformation of a four dimensional large $N$ CFT by such operators into the expression for the radial ADM Hamiltonian in the bulk is found. 

This prescription clarifies the role of various functions of background fields that appear in the flow equation defining the deformed holographic CFT, and also their relationship to the holographic anomaly.

The effect of these deformations can also be seen as triggering a generalized gradient flow for the fields of the induced gravity theory obtained from integrating out the fundamental fields of the holographic CFT. The potential for this gradient flow is found to resemble the two derivative effective action previously derived using holographic renormalization.

	\end{abstract}
\section{Introduction}
The holographic duality geometerizes the renormalization group flow of quantum field theories. In the cases where the bulk theory is classical general relativity coupled to some matter fields for example, the radial evolution in the bulk is equated to the renormalization group flow in the boundary theory. 

In the Wilsonian renormalization group, the scale associated to (effective) quantum field theories is identified with its cutoff. The beta functions and anomalous scaling dimensions of various operators can be deduced from the response of the quantum field theory to varying this cutoff. It is tempting to identify the radial coordinate with this cutoff scale and consider the families of holographic quantum field theories that arise from varying this scale as well. In  \cite{hemPol}, \cite{flrang} attempts were made to identify the radial cutoff in the bulk with the Wilsonian momentum cutoff on the field theory side. What seems however to work better to identify with the radial cutoff in the bulk, is a position space cutoff on the field theory side as emphasized in \cite{lecholrg}, \cite{SSLee}. 

In order to sharpen this intuition, a specific means of implementing a cutoff in holographic conformal field theory (CFT) which corresponds to putting a bulk cutoff at finite radial distance must be identified. The basic idea is that there are certain double trace deformations of the holographic conformal field theory which, in addition to certain functions of background fields, drive a flow on the space of quantum field theories that in the bulk is translated into the radial development generated by the normal component of the Einstein equations $G_{nn}$.
This in fact is the Arnowitt--Deser--Misner (or ADM) Hamiltonian of the theory (see \cite{ADM}) that one obtains from first foliating the bulk with hypersurfaces of constant radius and performing the Legendre transform.

Specifically, given the irrelevant nature of these deformations, they are driven to zero under the action of the renormalization group. In terms of the dual theory, this is to be interpreted as running in tandem with outwards radial development from a truncated AdS space of finite radius to one with an asymptotic boundary. 

A complementary way to view what is happening is by associating to the RG scale the energy of some probe in this truncated bulk spacetime. At high enough energies, such a probe would be sensitive to the cutoff surface or finite boundary. At lower energies however, as it probes the deep interior of this spacetime, the fact that the boundary is at finite radius becomes increasingly unimportant, and for a low enough energy probe, the boundary might as well be at infinity. This is a dual statement to the fact that the CFT, dual to the asymptotically AdS spacetime sits at the end of an RG flow that starts at the quantum field theory obtained from deforming said CFT by the aforementioned irrelevant operators.

In the context of pure gravity in the bulk of $AdS_{3}$, such a method was first proposed in \cite{mmv} and then it was pointed out that more double trace deformations are indeed needed to capture the dynamics of fields other than the metric in the bulk in 
\cite{lkrmar}. The effect of a related, irrelevant single trace deformation performing a similar role in the $AdS_{3}/CFT_{2}$ context was also considered in \cite{gir}, \cite{giv}. 
Then, the higher dimensional generalization of this deformation was considered in \cite{Nima}, and specialized to the context of holographic theories in \cite{mtay}. It was generalized in the holographic context further in \cite{hart} where the role of various background fields were also taken into account. 

The prescription of the authors of \cite{hart} is to define the effective field theory's generating functional through the following flow equation (in two, three or four dimensions):
\begin{equation}
\frac{\partial \textrm{ln}Z_{eff}}{\partial \lambda}= -\int \textrm{d}^{D}x \sqrt{g}X,\label{hfe}
\end{equation}
\begin{equation}
X=\left(T_{\mu\nu}+a_{D}r^{D-2}_{c}\tilde{C}_{\mu\nu}\right)^{2}-\frac{1}{D-1}\left(T^{\mu}_{\mu}+a_{d}r^{D-2}_{c}\tilde{C}^{\nu}_{\nu}\right)^{2}-\frac{r^{D}_{c}}{D\lambda}\left(\tilde{t}^{\mu}_{\mu}-\frac{\tilde{R}}{16\pi G}-a_{D}\tilde{C}^{\mu}_{\mu}\right).
\end{equation}
Here, $T_{\mu\nu}$ denotes the energy momentum tensor, the quantity $\tilde{C}_{\mu\nu}$ depends on the curvature tensors of the background geometry with metric $g_{\mu\nu}$. The constant $r_{c}$ denotes the radius in the bulk corresponding to the cutoff surface, and the constant $a_{D}$ stands for $\frac{1}{8\pi G(D-2)}$. The re-scaled Ricci scalar is denoted $\tilde{R}=r^{D-2}_{c}R$ and the tensor $\tilde{t}_{\mu\nu}$ is a function of sources other than the metric that are also turned on. 

A brief explanation for why the scale associated with these double trace deformations can also be seen as a cutoff in energy for the deformed holographic quantum field theory is necessary at this point. Following the authors of \cite{hart}, consider the theory defined on a simple geometry, namely a square torus of length $L$, and let us momentarily turn off sources other than the metric.\footnote{This can be done more generally, as the authors of \cite{hart} show, however, for the purpose of illustrating the role of $\lambda$ as a cutoff scale, this setting is sufficient} Also assume that there is no momentum carried in the state in which the energy momentum tensor's expectation values are evaluated. In this situation, the energy density is given by the expectation value of one of the components of the energy momentum tensor. This can be integrated up to find the energy $E(L,\lambda)$ of the system and it is found to be
\begin{equation}
E(L,\lambda)=\frac{(D-1)L^{D-1}}{2D\lambda}\left(1-\sqrt{1-\frac{4D\lambda E_{o}(L)}{(D-1)L^{D-1}}}\right),
\end{equation} 
where $E_{o}(L)$ denotes the energy of the underformed CFT. This function hits a square root singularity at $E^{*}=\frac{(D-1)L^{D-1}}{4D\lambda}$, after which the energies take imaginary values. The idea is to discard all energies above $E^{*}$ and treat this value of the energy as a cutoff. In fact, this equation could also be used to write $\lambda$ as $\frac{(D-1)L^{D-1}}{4DE^{*}}$ which clarifies its role as a cutoff scale. 

This expression arises from taking the radial bulk Hamiltonian constraint in AdS space and identifying the metric on a constant radius hypersurface in the bulk with $r^{2}_{c}g_{\mu\nu}$, the momentum conjugate to this metric with $\sqrt{g}((r_{c}^{2-D}T^{\mu\nu}-(D-1)g^{\mu\nu}(x,r=0))+a_{d}\tilde{C}^{\mu\nu})$, and the parameter $\lambda$ identified as
\begin{equation}
\lambda=\frac{4\pi G}{r_{c}D}.
\end{equation}
This allows us to rewrite the derivative $\frac{\partial}{\partial \lambda}$ in terms of the derivative $\frac{\partial}{\partial r_{c}}$ in the left hand side of \eqref{hfe}. 
For the full list of identifications, see \cite{hart}. The above equation is guaranteed to give rise to the right kind of bulk physics although it is unclear what the underlying coarse graining mechanism is that gives rise to such a flow. 

In what follows an alternative prescription to the one above is presented through which flow equations similar to the one above are derived for four dimensional, large $N$, holographic conformal field theories. The advantage of the approach presented here is that the functions of background fields appearing in \eqref{hfe} are shown to arise from certain cancellations rather than being posited in the definition of the effective field theory. 
The route taken is to start from the definition of a conformal field theory deformed by certain double trace operators:
\begin{equation}
\frac{\partial \textrm{ln}Z}{\partial \lambda}=\int \textrm{d}^{4}x \sqrt{g}\left(-\mu\langle TT\rangle-\frac{\kappa}{2}\langle \mathcal{O}\mathcal{O}\rangle\right) -\mathcal{A}[g,\phi],
\end{equation}
 where $g_{\mu\nu}(x)$ and $\phi(x)$ are sources for the operator $T^{\mu\nu}$ and $\mathcal{O}(x)$ respectively. The functional $\mathcal{A}[g,\phi]$ is the integrated conformal anomaly. 
 The reason why it appears here is that when the deformation parameters $\mu$ and $\kappa$ are set to zero, the flow equation reduces to the statement that the conformal field theory in the presence of arbitrary background fields is anomalous under scale transformations. 
 The scalar double trace deformation is formed from the single trace scalar operator $\mathcal{O}(x)$ which couples to source $\phi(x)$, and its expectation value is denoted by $\langle \mathcal{O}\mathcal{O}\rangle$. Similarly, $\langle TT\rangle$ denotes the expectation value of the following operator:
 \begin{equation}
 TT(x)\equiv \left(G_{\mu\nu\rho\sigma}T^{\mu\nu}T^{\rho\sigma}\right)(x), \label{tay}
 \end{equation}
 where $G_{\mu\nu\rho\sigma}=g_{\mu(\rho}g_{\sigma)\nu}-\frac{1}{3}g_{\mu\nu}g_{\rho\sigma}$ is the de Witt supermetric in four dimensions.\footnote{I will assume tentatively that an appropriate regularization prescription has been chosen to define these double trace operators but this will not feature in anything to follow because I will be using large $N$ factorization to make sense of the expectation values of the operators in the above flow equation.}
 
 The operator \eqref{tay} is the same as the one proposed by Taylor in \cite{mtay} as the higher dimensional generalizations of the two dimensional $T\bar{T}$ deformation introduced in \cite{fzam}, \cite{tat}. 
 
 In the large $N$ limit, the expectation value of the double trace operators factorizes:
 \begin{equation}
 \langle TT\rangle=G_{\mu\nu\rho\sigma}\langle T^{\mu\nu}\rangle \langle T^{\rho\sigma}\rangle, \ \ \langle \mathcal{O}\mathcal{O}\rangle=\langle \mathcal{O}\rangle^{2},
 \end{equation}
 and so the equation that we will use reads 
 \begin{equation}
 \frac{\partial \textrm{ln}Z[g,\phi]}{\partial \lambda}=-\int \textrm{d}^{4}x \sqrt{g}\left(\mu G_{\mu\nu\rho\sigma}\langle T^{\mu\nu}\rangle\langle T^{\rho\sigma}\rangle+\frac{\kappa}{2} \langle\mathcal{O}\rangle^{2}\right)-\mathcal{A}[g,\phi]. \label{fe}
 \end{equation}
The left hand side can be rewritten as follows 
\begin{equation}
\frac{\partial \textrm{ln}Z[g,\phi]}{\partial \lambda}=\int \textrm{d}^{4}x \sqrt{g}\left(\beta_{\mu\nu}(g,\phi)\frac{\delta }{\delta g_{\mu\nu}}+\beta_{\phi}(g,\phi)\frac{\delta}{\delta \phi}\right)\textrm{ln}Z[g,\phi]. \label{fle}
\end{equation}
where the flow functions are defined as:
\begin{equation}
\beta_{\mu\nu}(g,\phi)=\frac{\partial g_{\mu\nu}}{\partial \lambda}, \ \ \beta_{\phi}(g,\phi)=\frac{\partial \phi}{\partial \lambda}. \label{flf}
\end{equation}
Similar flow functions are defined in \cite{lecholrg} although the beta functions there also depend on the expectation values of the deforming operators, and lead in a different manner to RG flow equations that can be mapped into the bulk Hamiltonian and momentum constraints.

Going back to our flow equation \eqref{fle}, the exercise now is to find an appropriate change of variables so that the equation 
\begin{equation*}
\int \textrm{d}^{4}x \sqrt{g} \left(\beta_{\mu\nu}(g,\phi)\frac{\delta}{\delta g_{\mu\nu}}+\beta_{\phi}(g,\phi)\frac{\delta }{\delta \phi}\right)\textrm{ln}Z[g,\phi]=
\end{equation*}
\begin{equation}
-\int \textrm{d}^{4}x \sqrt{g}\left(\mu G_{\mu\nu\rho\sigma}\langle T^{\mu\nu}\rangle\langle T^{\rho\sigma}\rangle+\frac{\kappa}{2} \langle\mathcal{O}\rangle^{2}\right)-\mathcal{A}[g,\phi], \label{ffe}
\end{equation}
can be rewritten as the radial Hamiltonian of a gravitational theory in one higher dimension. In doing so, we will find that the flow functions \eqref{flf} must take a very particular form. 
\section{From the deformation equation to the bulk Hamiltonian}
The response of the quantum field theory of interest under the infinitesimal change of scale associated to the addition of the scalar and stress tensor double trace deformations is expressed by \eqref{ffe}. The change in $\lambda$ should translate into the change in the radial cutoff in the bulk, which strictly speaking translates into a normal deformation of an embedded constant radius hypersurface, namely the boundary. Such transformations, on the phase space of general relativity coupled to scalar matter say, are generated by constraint equations. 

The phase space is parameterized by $(g_{\mu\nu}(x,r),\phi(x,r),\pi^{\mu\nu}(x,r),p_{\phi}(x,r))$, i.e. the metric on the hypersurface and its conjugate momentum in addition to the value of a scalar field on the hypersurface and its momentum, and the constraints that compose the ADM Hamiltonian of general relativity read:
\begin{eqnarray}
H(N)=\int \textrm{d}^{4}x N(x)\left\{ \frac{-1}{\sqrt{g}}\left(G_{\mu\nu\rho\sigma}\pi^{\mu\nu}\pi^{\rho\sigma}+p^{2}_{\phi}\right)-\sqrt{g}\left(-\frac{(\partial^{\mu}\phi\partial_{\mu}\phi)}{2}+R+V(\phi)\right) \right\}=0,\label{hc} \\
H_{\mu}(\xi^{\mu})=\int \textrm{d}^{4}x \xi^{\mu}(x)\left(\nabla_{\nu}\pi^{\mu\nu}+p_{\phi}\nabla^{\mu}\phi\right)=0,\label{vc} \\
H_{tot}=H(N)+H_{\mu}(\xi^{\mu}).
\end{eqnarray}
Here, $(N,\xi^{\mu})$ denote lagrange multipliers for these constraints. 

The scalar function \eqref{hc} is the constraint associated to normal deformations of a constant radius hypersurface, i.e. the Poisson brackets of phase space functions with this constraint yields the transformation law for said functions under hypersurface orthogonal deformations generated by a diffeomorphism of the ambient spacetime (on shell). Similarly, the vector constraint \eqref{vc} generates through its Poisson brackets with functions on phase space the deformations tangential to the hypersurface, or in other words, the diffeomorphisms tangential to the hypersruface. 

The fact that these constraints generate the deformations of hypersurfaces generated by a diffeomorphism of the ambient spacetime is encoded in the Poisson algebra of the constraints. This algebra reads as follows:
\begin{eqnarray}
\left\{H(N),H(M)\right\}=H_{\mu}(g^{\mu\nu}(N\partial_{\nu}M-M\partial_{\nu}N)), \\ \label{diralg1}
\left\{H(N),H_{\nu}(\xi^{\nu})\right\}=-H(\xi^{\mu}\partial_{\mu}N),\\ \label{diralg2}
\left\{H_{\mu}(\xi^{\mu}),H_{\nu}(\zeta^{\nu})\right\}=H_{\mu}([\xi,\zeta]^{\mu}). \label{diralg3}
\end{eqnarray}
Here, the bracket $[,]$ is the Lie bracket of vector fields on the hypersurface.

These relations reflect the commutator algebra (as pointed out in \cite{bunster}) obtained from decomposing a spacetime vector field $v^{A}$ (where the capital latin index runs over the $D+1$ dimensions of spacetime) into components normal and tangential to the constant radius hypersurface: $v^{A}=v_{n}n^{A}+v^{A}_{\parallel}$,  where $n^{A}$ is the normal to the hypersurface and the vector $v^{A}_{\parallel}$ has indices only tangential to the hypersurface. Then, the action of the diffeomorphism transformation is generated by the constraints smeared as $(H(v_{n}), H_{\mu}(v^{\mu}_{\parallel}))$. 

Thus, the flow equation \eqref{ffe} should be translated into the scalar constraint \eqref{hc} through an appropriate change of variables. The aim of this section is to find this change of variables. 

It will help to start by turning off the scalar deformation sourced by $\phi$
and consider just the stress tensor sector of the theory, which in the bulk corresponds to pure gravity. 

\subsection{Pure gravity in the bulk}
In this case, the flow equation \eqref{ffe} reads 
\begin{equation*}
\int \textrm{d}^{4}x \sqrt{g}\beta_{\mu\nu}(g)\frac{\delta \textrm{ln}Z[g]}{\delta g_{\mu\nu}}=\int \textrm{d}^{4}x \sqrt{g}\sigma(x)\beta_{\mu\nu}(g)\langle T^{\mu\nu}\rangle =
\end{equation*}
\begin{equation}
=\int \textrm{d}^{4}x \sqrt{g}\left(-\mu G_{\mu\nu\rho\sigma}\langle T^{\mu\nu}\rangle \langle T^{\rho\sigma} \rangle \right)-\mathcal{A}[g].
\end{equation}
The aim is to convert this into the radial Hamiltonian constraint of general relativity operating in one higher dimension.
The idea is to identify the theory's background metric with the metric induced on a constant radius hypersurface and relate the one point function of the energy momentum tensor in a general background to the momentum conjugate to this metric. 

The latter is what we will pay attention to here. The discrepency between the energy momentum tensor's one point function and the momentum is captured by the ambiguity in adding local counterterms. By counterterms here, I am referring to local functionals of the source, here the metric, and its derivatives that are added to the generating functional. These aren't added in order to cancel any divergences although in the limit where the regularization is removed, this would be their purpose.
 In other words:
\begin{equation}
\sqrt{g}\langle T^{\mu\nu}\rangle=\pi^{\mu\nu}+\frac{\delta S[g]}{\delta g_{\mu\nu}}.
\end{equation}
Plugging this into the flow equation leads to the following equation:
\begin{equation*}
\int \textrm{d}^{4}x \sqrt{g}\left(\beta_{\mu\nu}\pi^{\mu\nu}+\beta_{\mu\nu}\frac{\delta S[g]}{\delta g_{\mu\nu}}\right)=\end{equation*}
\begin{equation}=\int \textrm{d}^{4}x \sqrt{g} \left(-\frac{\mu}{\sqrt{g}} G_{\mu\nu\rho\sigma}\pi^{\mu\nu}\pi^{\rho\sigma}+2\mu G_{\mu\nu\rho\sigma}\pi^{\mu\nu}\frac{\delta S[g]}{\delta g_{\mu\nu}}-\mu G_{\mu\nu\rho\sigma}\frac{\delta S[g]}{\delta g_{\mu\nu}}\frac{\delta S[g]}{\delta g_{\rho\sigma}}\right)-\mathcal{A}(g). 
\end{equation}
First, we require the cancellation of the term linear in the momentum. This gives us the following gradient condition:
\begin{equation}
\beta_{\mu\nu}(g)=2\mu G_{\mu\nu\rho\sigma}\frac{\delta S[g]}{\delta g_{\rho\sigma}}. 
\label{gfe}
\end{equation}
The first term on the left hand side cancels against the second term on the right hand side. Then, we also see that the second term on the left hand side reads
\begin{equation}
\beta_{\mu\nu}\frac{\delta S[g]}{\delta g_{\mu\nu}}=2\mu G_{\mu\nu\rho\sigma}\frac{\delta S[g]}{\delta g_{\mu\nu}}\frac{\delta S[g]}{\delta g_{\rho\sigma}}.
\end{equation}
And, the flow equation then reduces to 
\begin{equation}
\int \textrm{d}^{4}x \sqrt{g}\left(-\frac{\mu}{\sqrt{g}}G_{\mu\nu\rho\sigma}\pi^{\mu\nu}\pi^{\rho\sigma}+\mu G_{\mu\nu\rho\sigma}\frac{\delta S[g]}{\delta g_{\mu\nu}}\frac{\delta S[g]}{\delta g_{\rho\sigma}}\right)-\mathcal{A}(g)=0. \label{alhc}
\end{equation}

The  integrated Weyl anomaly for four dimensional conformal field theories takes the form:
\begin{equation}
\mathcal{A}[g]=\int \textrm{d}^{4}x \sqrt{g}\left(\frac{c}{3}-a\right)R^{2}+(-2c+4a)R^{\mu\nu}R_{\mu\nu}+(c-a)R^{\mu\nu\rho\sigma}R_{\mu\nu\rho\sigma},
\end{equation}

Then, recalling that the theory we are dealing with is holographically dual to general relativity, requires $a=c$, as discovered in \cite{Skenderis}. So, we have
\begin{equation}\mathcal{A}[g]=\int \textrm{d}^{4}x \sqrt{g}2a\left(R^{\mu\nu}R_{\mu\nu}-\frac{1}{3}R^{2}\right)\equiv a A^{(a=c)}[g].\end{equation}

For the function $S[g]$, if we take the Holographic counterterm of \cite{BK}, i.e.\footnote{Again, this is just a local function that has the same form as the Holographic couterterm, but it isn't added with the intention to cancel any divergences, because at a finite radius boundary, there aren't any.}
\begin{equation}
S[g]=\frac{3}{\ell}\left(\int \textrm{d}^{4}x\sqrt{g}\left(1+\frac{\ell^{2}R}{6}\right)\right),
\end{equation}
then we find that first, the anomaly term cancels against the following term:
\begin{equation}
A^{(a=c)}=\int \textrm{d}^{4}x \sqrt{g}G_{\mu\nu\rho\sigma}\frac{\delta\left(\int \textrm{d}^{4}x \sqrt{g}R \right)}{\delta g_{\mu\nu}}\frac{\delta\left(\int \textrm{d}^{4}x \sqrt{g}R\right)}{\delta g_{\rho\sigma}},
\end{equation}
if the identification $\frac{\mu\ell^{2}}{4}=a$ is made. This form of the $a=c$ anomaly also appeared in \cite{dilax}.
Recalling that $a=\frac{\pi^{2} \ell^{3}}{8 G}$, for supergravity in AdS$_{5}\times S^{5}$ which is dual to  $\mathcal{N}=4$ Super Yang-Mills theory at large $N$, the relation found here implies that $\mu=\frac{\pi^{2} \ell}{2G}$. 

A caveat is in order at this point: note that the correct counter-term to use at asymptotic infinity is the one presented in \cite{counterright} that involves in addition to what is above a logarithmically divergent (as the asymptotic boundary is approached) term that multiplies a fourth derivative term arising from the anomaly itself. Thus an additional canonical transformation which involves the addition of that counter term must be performed before taking the asymptotic limit of quantities defined in the setup considered here in order to avoid divergences that would otherwise arise in the action on shell. 

Then, the remaining terms in \eqref{alhc} organise themselves into 
\begin{equation}
H=-\int \textrm{d}^{4}x \left(\frac{G_{\mu\nu\rho\sigma}\pi^{\mu\nu}\pi^{\rho\sigma}}{\sqrt{g}}-\sqrt{g}\left(R+\frac{12}{\ell^{2}}\right)\right)=0.
\end{equation}
This is nothing but the Hamiltonian constraint of five dimensional general relativity with a negative cosmological constant, smeared against unit lapse. 
We also have the vector constraint density:
\begin{equation}
\nabla_{\mu}\pi^{\mu}_{\nu}=0,
\end{equation}
that follows from the covariant conservation of the stress tensor, and can be integrated against an arbitrary shift vector field to obtain the constraint.

The flow function now reveals that the boundary metric satisfies a generalized Ricci flow equation:
\begin{equation}
\beta_{\mu\nu}(g)=\mu \left(\frac{1}{\ell}g_{\mu\nu}+\frac{\ell}{2}\left(R_{\mu\nu}-\frac{1}{6}R g_{\mu\nu}\right)\right).
\end{equation}
This is similar to what was found in \cite{grg} and \cite{Nakayama1}.

\subsection{Including the scalar matter}
In this section, the effect of adding a double trace scalar operator deformation to the above setup is considered. The effect of adding this deformation in the bulk is similar to that of the stress tensor double trace deformation in that it regulates the theory on the boundary in a way that corresponds to pulling the boundary to a finite radius. 

First, note that the source $\phi(x)$ of the single trace scalar deformation is also space dependent and hence it contributes towards the anomaly, and for holographic theories it reads:\footnote{See \cite{Osborn1} the form of the scalar source and metric dependent anomaly in general CFTs}
\begin{equation*}
\mathcal{A}[g,\phi]=\int \textrm{d}^{4}x \sqrt{g}\bigg(\alpha(\phi)R^{\mu\nu}R_{\mu\nu}-\gamma(\phi)R^{2}+\zeta(\phi)\partial_{\mu}\phi\partial^{\mu}R+\eta(\phi)R(\partial^{\mu}\phi \partial_{\mu}\phi)+\end{equation*}
\begin{equation}
+\xi(\phi)\partial^{\mu}\phi\partial^{\nu}\phi\left(R_{\mu\nu}-\frac{1}{2}Rg_{\mu\nu}\right)+\lambda(\phi)\nabla^{2}\phi
\nabla^{2}\phi+\chi(g,\phi)_{\mu\nu\rho\sigma}\partial^{\mu}\phi\partial^{\nu}\phi\partial^{\rho}\phi\partial^{\sigma}\phi\bigg).	\end{equation}

Then the various functions are found in terms of those that go into the definition of the function $S[g,\phi]$ that specifies the canonical transformation relating the one point function of $\mathcal{O}$ to the momentum $p_{\phi}$ conjugate to the source $\phi$ in the bulk Hamiltonian. 

Recall that the flow equation now looks like:
\begin{equation*}
-\int \textrm{d}^{4}x \sqrt{g}\left(\beta_{\mu\nu}(g,\phi)\frac{\delta}{\delta g_{\mu\nu}}+\beta_{\phi}(g,\phi)\frac{\delta}{\delta \phi}\right)\textrm{ln}Z[g,\phi]=
\end{equation*}
\begin{equation}
= -\int \textrm{d}^{4}x \sqrt{g}\left( \mu G_{\mu\nu\rho\sigma}\langle T^{\mu\nu}\rangle \langle T^{\rho\sigma}\rangle +\frac{\kappa}{2}\langle \mathcal{O}\rangle^{2}\right)+\mathcal{A}[g,\phi].
\end{equation}
Like in the pure gravity case, we make the identification
\begin{equation}
\sqrt{g}\langle \mathcal{O}\rangle=p_{\phi}-\frac{\delta S[g,\phi]}{\delta \phi},
\end{equation}
in addition to the identification made of the momentum conjugate to the metric made in the previous section. 

The flow equation then becomes
\begin{equation*}
\int \textrm{d}^{4}x \left(\beta_{\mu\nu}(g,\phi)\left(\pi^{\mu\nu}-\frac{\delta S[g,\phi]}{\delta g_{\mu\nu}}\right)+\beta_{\phi}(g,\phi)\left(p_{\phi}-\frac{\delta S[g,\phi]}{\delta \phi}\right)\right)=
\end{equation*}
\begin{equation}
\int \textrm{d}^{4}x\left(-\mu\frac{G_{\mu\nu\rho\sigma}}{\sqrt{g}}\left(\pi^{\mu\nu}-\frac{\delta S[g,\phi]}{\delta g_{\mu\nu}}\right)\left(\pi^{\rho\sigma}-\frac{\delta S[g,\phi]}{\delta g_{\rho\sigma}}\right)-\frac{\kappa}{2\sqrt{g}}\left(p_{\phi}-\frac{\delta S[g,\phi]}{\delta \phi}\right)^{2}\right)-\mathcal{A}[g,\phi]. 
\end{equation}
then, like in the gravity case, we set: 
\begin{equation} 
\beta_{\phi}(g,\phi)=\kappa \frac{\delta S[g,\phi]}{\delta \phi}. \label{sfe}
\end{equation}
Then, the final expression for the bulk Hamiltonian reads 
\begin{equation*}
\int\textrm{d}^{4}x \left(-\mu\frac{G_{\mu\nu\rho\sigma}}{\sqrt{g}}\pi^{\mu\nu}\pi^{\rho\sigma}-\kappa\frac{p^{2}_{\phi}}{2\sqrt{g}}\right)=\end{equation*}\begin{equation}-\mathcal{A}[g,\phi]+\int \textrm{d}^{4}x \left(\mu G_{\mu\nu\rho\sigma}\frac{\delta S[g,\phi]}{\delta g_{\mu\nu}}\frac{\delta S[g,\phi]}{\delta g_{\rho\sigma}}+\frac{\kappa}{2}\left(\frac{\delta S[g,\phi]}{\delta \phi}\right)^{2}\right). \label{prehc}
\end{equation}
Given that the scalar field (like any matter field) couples to gravity with a strength set by Newton's constant which has been set to unity, and that there is no other scale involved in this coupling requires $\mu=\kappa$. Then we divide throughout by $\mu$ and absorb the remaining factor of it in the denominator of the term involving $\mathcal{A}[g,\phi]$ by redefining the as yet undetermined functions in its definition. This equality is significant, as it implies that there is just one scale associated to the total double trace deformation. More will be said about this condition in the discussion. 

In order for the momentum independent part of the above constraint equation to be of the form $\sqrt{g}\left(-\frac{(\partial_{\mu}\phi\partial^{\mu}\phi)}{2}+R+V(\phi)\right)$, the potential $S[g,\phi]$ can have two derivatives of the sources at most. 

The ansatz made is 
\begin{equation}
S[g,\phi]=\int \textrm{d}^{4}x \sqrt{g}\left(X(\phi)+U(\phi)R+P(\phi)(\partial^{\mu}\phi\partial_{\mu}\phi)\right). \label{pot}
\end{equation}
If we require the cancellation between the anomaly and the square of the derivatives of this functional present in the momentum independent terms of \eqref{prehc}, then the flow equation now reads as the Hamiltonian constraint for the five dimensional bulk gravity- scalar system:
\begin{equation}
\int \textrm{d}^{4}x \left\{
-\frac{1}{\sqrt{g}}\left(G_{\mu\nu\rho\sigma}\pi^{\mu\nu}\pi^{\rho\sigma}+\frac{p^{2}_{\phi}}{2}\right)-\sqrt{g}\left(\frac{1}{2}\partial_{\mu} \phi\partial^{\mu}\phi+R+V(\phi)\right)\right\}=0,
\end{equation}
provided the following superpotential like relations are satisfied:
\begin{eqnarray}
\left(\frac{X'^{2}}{2}-\frac{X^{2}}{3}\right)=V,\ \ \label{suppot1}
\left(U'X'-\frac{UX}{3}\right)=1, \ \ \label{suppot2}
\left(PX'\right)'-PX=1. \label{suppot3}
\end{eqnarray}
As promised before, the functions in the anomaly are related to the functions in the defintion of $S$, i.e.
\begin{eqnarray}
\chi(g,\phi)_{\mu\nu\rho\sigma}=P^{2}\left(g_{\mu(\rho}g_{\sigma)\nu}-g_{\mu\nu}g_{\rho\sigma}\right),\alpha(\phi)=\frac{U^{2}}{2} \ \  \\ \label{anomfun1}
\gamma(\phi)=\left(\frac{U^{2}}{3}-\frac{U'^{2}}{2}\right),\ \ \zeta(\phi)=PU' \\ \label{anomfun2}
\eta(\phi)=(PU')', \ \ \xi(\phi)=PU,\ \  \lambda(\phi)=\frac{P^{2}}{2}.  \label{anomfun3}
\end{eqnarray}
This analysis reproduces a part of the results obtained in \cite{dilax} where a completely generic scalar coupled to AdS gravity was first considered. 
In all these expressions, the prime denotes partial differentiation with respect to $\phi$, i.e. $()'\equiv\partial_{\phi}()$. Note that in deriving these conditions, many integrations by parts have been carried out and boundary terms have been discarded. 

Some of these relations are identical to those derived previously in the context of holographic renormalization (in say, \cite{twoder}).
The flow function for the scalar field can also be interpreted as the beta function for the renormalization group flow triggered by the addition of the deformation. To leading order in perturbation theory, the beta function takes the form
\begin{equation}
\beta_{\phi}(g,\phi)\propto (4-\Delta)\phi+\cdots,
\end{equation}
where $\Delta$ is the conformal dimension of the operator $\mathcal{O}$. This implies that the function $X(\phi)$ is given to leading order by:
\begin{equation}
X(\phi)=-\frac{6}{\ell}-\frac{1}{2\ell}(4-\Delta)\phi^{2}+\cdots
\end{equation}
The reason for the specific numerical factor in the $\phi$ independent part and the factors of $\ell$ being where they are is to ensure that the bulk potential computed through the relation:
\begin{equation}
V=\frac{X'^{2}}{2}-\frac{X^{2}}{3}=\frac{12}{\ell^{2}}-\frac{1}{2}\frac{\Delta(\Delta-4)\phi^{2}}{\ell^{2}}+\cdots
\end{equation}
becomes the appropriate cosmological constant in the pure gravity Hamiltonian constraint when $\phi\rightarrow 0$. 

Also, note that the mass-conformal dimension relationship
\begin{equation}
m^{2}=\frac{\Delta(\Delta-4)}{\ell^{2}},
\end{equation}
has been recovered. So, for example, when $\Delta=4$, i.e. when $\mathcal{O}$ is marginal, the bulk scalar field is massless and minimally coupled. 

The expression of diffeomorphism invariance tangential to the hypersurfaces in this coupled system, takes the form:
\begin{equation}
\nabla_{\nu}\pi^{\mu\nu}+p_{\phi}\nabla^{\mu}\phi=0,\label{vecc}
\end{equation}
which is the vector constraint in the bulk, whose form follows again from the Ward identity associated to diffeomorphism invariance of the holographic theory coupled to the metric and scalar sources.
\subsection{Closure and Cancellation}
Now we see that the key feature of the holographic anomaly which allows us to convert the flow equation \eqref{ffe} into the constraint equation \eqref{hc} is the following property:
\begin{equation*}
\mathcal{A}[g,\phi]-\int \textrm{d}^{4}x\left(G_{\mu\nu\rho\sigma}\frac{\delta S[g,\phi]}{\delta g_{\mu\nu}}\frac{\delta S[g,\phi]}{\delta g_{\rho\sigma}}+\frac{1}{2}\left(\frac{\delta S[g,\phi]}{\delta \phi}\right)^{2}\right)=\end{equation*}\begin{equation}=\int\textrm{d}^{4}x \sqrt{g}\left(-\frac{1}{2}\left(\partial^{\mu}\phi\partial_{\mu}\phi\right)+R+V\right).\label{cencel}
\end{equation}
This holds, provided the relations \eqref{suppot1}-\eqref{anomfun3} all hold. In fact, these relations were derived above by requiring \eqref{cencel} to hold, which in turn came from wanting to transform the flow equation \eqref{ffe} into the constraint \eqref{hc}. 

We could however have chosen to derive this relation from an alternative, yet equivalent demand. Namely, starting from \eqref{ffe} and making the appropriate change of variables, we are led to \eqref{prehc}, which I reproduce here:
\begin{equation*}
H=\int \textrm{d}^{4}x N(x)\bigg\{ -\mu\frac{G_{\mu\nu\rho\sigma}}{\sqrt{g}}\pi^{\mu\nu}\pi^{\rho\sigma}-\kappa\frac{p^{2}_{\phi}}{2\sqrt{g}}+\end{equation*}\begin{equation}+\left(\mathcal{A}[g,\phi]-\mu G_{\mu\nu\rho\sigma}\frac{\delta S[g,\phi]}{\delta g_{\mu\nu}}\frac{\delta S[g,\phi]}{\delta g_{\rho\sigma}}-\frac{\kappa}{2}\left(\frac{\delta S[g,\phi]}{\delta \phi}\right)^{2}\right)\bigg\}=0.
\end{equation}
Note that independently of this equation, we are granted the vector constraint \eqref{vecc}, which simply follows form the diffeomorphism Ward identity of the energy momentum tensor and the scalar source. 

Without assuming that the final expression for the scalar constraint should take any particular form, but simply that its Poisson algebra closes, i.e. that 
\begin{equation}
\left\{H(N),H(M)\right\}\approx 0, \label{os}
\end{equation}
where the symbol $\approx$ here denotes equality on the sub-space of phase space where the constraints are satisfied \footnote{The right hand side of \eqref{os} is in general some combination of the constraints $H(N)$ and the vector constraint, and so it vanishes when they are satisfied. In the language of constrained Hamiltonian dynamics, satisfying this closure property qualifies constraints to be `first class'.}, implies that the relation \eqref{cencel} must hold. This is guaranteed by a theorem proven in \cite{farkas} and strengthened further in \cite{me2}. The assumptions that go behind the theorem are simply that the vector constraint holds and that the momentum dependent part of the scalar constraint is quadratic and ultralocal. Although it was proven there for the pure gravity case, it is not hard to see that for the case at hand, where the scalar field's kinetic term in the Hamiltonian takes the form $\frac{p^{2}_{\phi}}{2\sqrt{g}}$, the potential term has to contain a derivative independent potential $V(\phi)$ in addition to the term $-\frac{(\partial_{\mu}\phi\partial^{\mu}\phi)}{2}$ in order to close at all. 

Note that the closure condition \eqref{os} is in fact weaker than the statement of the Poisson algebra \eqref{diralg1}-\eqref{diralg3}. There, the specific combination of constraints that vanish on the right hand side of \eqref{os} and the corresponding structure functions (as opposed to constants due to the field dependence) are also known. Thus, the condition \eqref{os} is not quite the same as demanding the emergence of bulk diffeomorphism invariance which is what is encoded in \eqref{diralg1}-\eqref{diralg3}. It is the non-linear generalization of the requirement that five dimensional general relativity only propagate the spin two modes of the graviton in addition to the coupled scalar field. 

That being said, the uniqueness of the form of the constraint functions that satisfy \eqref{os} given \eqref{vecc} implies that the only way to satisfy the closure condition is in the manner general relativity does, i.e. \eqref{diralg1}. In other words, there is something unique about how the diffeomorphism invariance tangential to an embedded hypersurface gets promoted to the full diffeomorphism invariance of the ambient spacetime, at least at the level of the constraints that generate the corresponding transformations on phase space. 

The requirement that the momentum dependent `kinetic term' in the constraint equation be quadratic and ultralocal, i.e. of the form
\begin{equation}
-\frac{G_{\mu\nu\rho\sigma}\pi^{\mu\nu}\pi^{\rho\sigma}}{\sqrt{g}}-\frac{p^{2}_{\phi}}{\sqrt{g}}, 
\end{equation}
is equivalent to the statement that the radial velocities in the bulk of the fields $g_{\mu\nu}$ and $\phi$ are at most linear in the momenta. In other words:
\begin{equation}
\sqrt{g}\left(\beta_{\mu\nu}(g,\phi)-\nabla_{(\mu}\xi_{\nu)}\right)=\pi_{\mu\nu}-\frac{1}{2}g_{\mu\nu}\pi^{\rho}_{\rho},\ \
\sqrt{g}(\beta_{\phi}(g,\phi)-\xi^{\nu}\nabla_{\nu}\phi)=p_{\phi}. 
\end{equation}

This feature of the flow was also noted in \cite{hart}, and follows directly from the structure of the double trace deformations being added, and is not sensitive to the background fields organising themselves into any particular form, which is what \eqref{cencel} reflects. We see here however that these two features are intimately related through the closure condition.

In the rest of this article, $\beta_{\phi}(g,\phi), \beta_{\mu\nu}(g,\phi)$ are interpreted not just as beta functions for the holographic quantum field theory, but also as the flow functions of the gradient flow regularization applied to the induced gravity theory obtained from integrating out the fields of this quantum field theory.  
\section{Generalized gradient flows and holography}
One way to think of the holographic duality is that it operates between quantum field theories in $D$ dimensions and gravitational theories in $D+1$ dimensions. However, when arbitrary background sources are turned on in the quantum field theory, there is an alternative statement of the duality which is seemingly more mundane, but perhaps still illuminating. First, one interprets the generating functional of the quantum field theory in the presence of said sources, metric included, as a non-local induced gravity theory. Then, the holographic duality implies that this $D$ dimensional non-local induced gravity theory and the $D+1$ dimensional theory, i.e. general relativity coupled to matter fields with negative cosmological constant, are equivalent to each other. This way of thinking about the duality was emphasised in \cite{tseytliu}. 

This section aims to explain how this duality manifests when the boundary is at finite radius. The key mechanism behind this version of the duality will be the generalized gradient flow, that will be discussed briefly in what folllows. 

The gradient flow (\cite{wilfow}, \cite{aokisan0}) is a method to regulate the correlation functions of composite operators of various quantum field theories in the coincidence limit. The idea is to append an additional dimension to the space on which the quantum field theory lives, and declare that the dependence of the fields of the theory on this additional dimension are dictated by gradient flow conditions. It was introduced in the context of Yang-Mills theory but can be applied to more general quantum field theories as well.
\subsection{How gradient flows lead to smearing}
To illustrate the idea, consider the simple example of applying it to the $O(N)$ non-linear sigma model in two dimensions.\footnote{The explanations here closely follow those in \cite{aokisan0}}. The action for the theory reads 
\begin{equation}
S[\sigma]=\frac{1}{2g^{2}}\int \textrm{d}^{2}x \left(h_{ab}(\sigma)\left(\partial_{\mu}\sigma^{a}\partial^{\mu}\sigma^{b}\right)\right), 
\end{equation}
where the fields $\sigma^{a}(x)$ are multi component scalars and the metric $h_{ab}(\sigma)$ is given by the inverse of
\begin{equation}
h^{ab}(\sigma)=\delta^{ab}-\sigma^{a}\sigma^{b}.
\end{equation}
The generalized gradient flow method for regularizing the divergences in correlators of composite operators in the contact limit starts by appending an additional dimension to the two dimensional space on which the field theory lives with an additional scale dimension $\lambda$:
$\sigma^{a}(x)\rightarrow \sigma^{a}(x,\lambda)$. Then, the dependence of the fields along this dimension is dictated by the flow equation
\begin{equation}
\frac{\partial \sigma^{a}(x,\lambda)}{\partial \lambda}=-h^{ab}(\sigma(x,\lambda))\frac{\delta S[\sigma(x,\lambda)]}{\delta \sigma^{b}}|_{\sigma^{a}(x)\rightarrow\sigma^{a}(x,\lambda)}.
\end{equation}
The $|_{\sigma^{a}(x)\rightarrow\sigma^{a}(x,\lambda)}$ denotes the boundary condition $\sigma^{a}(x,\lambda=0)=\sigma^{a}(x)$ at the boundary of the three dimensional space. The potential driving this flow, $S[\sigma(x,\lambda)]$ is the non-linear sigma model action where the fields $\sigma^{a}(x)$ are extended to the `flowed' fields $\sigma^{a}(x,\lambda)$. 

Explicitly, the flow equation reads:
\begin{equation}
\frac{\partial \sigma^{a}(x,\lambda)}{\partial \lambda}=\partial_{\mu}\partial^{\mu}\sigma^{a}+\sigma^{a}(\partial_{\mu}\sigma^{b} \partial^{\mu}\sigma_{b})+\frac{\sigma^{a}(\partial^{\mu}\sigma^{b})(\partial_{\mu}\sigma_{b})}{4(1-\sigma^{c}\sigma_{c})},
\end{equation}
where the indices of the internal vector components are contracted with the metric $h_{ab}(\sigma)$. For simplicity, consider the linearization of the above equation. This takes the form of a heat equation with the role of time being played by the flow parameter $\lambda$. The solution to the linearized equation then implies that to leading order in $\lambda$, the dependence of $\sigma^{a}(x,\lambda)$ on the flow time is given by smearing the original fields $\sigma^{a}(x)$ with the heat kernel: 
\begin{equation}
\sigma^{a}(x,\lambda)=\exp{\left(\lambda\left(\partial^{2}\right)\right)}\sigma^{a}(x)+\cdots
\end{equation}
The non locality associated to this smearing is what renders the correlators of composite operators built out of $\sigma^{a}(x,\lambda)$ finite. 
\subsection{Gradient flows for the induced boundary gravity theory}

The similarity between this general procedure and the holographic duality leads one to wonder whether they coincide when applied to theories expected to possess holographic duals. This was the focus of \cite{floweq1}, \cite{floweq2}, \cite{aokisan1}, \cite{aokisan2} and \cite{aokisan3}, where the generalized gradient flow is applied to the $O(N)$ vector model, and various aspects of the holographic duality having to do with reconstructing the bulk metric, understanding the effects of diffeomorphisms in the bulk and even computing $1/N$ corrections to the cosmological constant were considered. 

Here the aim is more modest, unlike the work mentioned in the previous paragraph which aims to understand the duality constructively through implementing the gradient flow, here we just notice there are gradient flows hiding in finite radius holography as implemented through the double trace deformations described in the previous sections. First, it would help to identify which theory this procedure is being applied to. The equations 
\begin{equation}
\frac{1}{\mu}\frac{\partial g_{\mu\nu}}{\partial \lambda}=G_{\mu\nu\rho\sigma}\frac{\delta S[g,\phi]}{\delta g_{\rho\sigma}}, \ \ \frac{1}{\mu}\frac{\partial \phi}{\partial \lambda}=\frac{\delta S[\phi,g]}{\delta \phi},
\end{equation}
certainly seem to have the structure of the generalized gradient flow equations barring the fact that they are not describing the flow of the fundamental fields of the theory itself, but of its sources. This motivated Jackson et. al. in \cite{grg} to call equations such as these `geometric RG flow' equations. However, if one considers the induced gravity theory in four dimensions that arises from integrating out the fundamental fields of the CFT. It is obtained by computing the generating functional as a function of the sources:
\begin{equation}
\textrm{ln}Z[g,\phi]=\mathcal{S}^{ind}[g,\phi],
\end{equation}
and interpret the resulting, non-local function of the metric and the other sources as the effective action for a four dimensional gravitational theory. The holographic duality appears to really coincide with applying the gradient flow regularization to this induced gravity theory, where the dependence of its fields $g_{\mu\nu}$ and $\phi$ on the additional dimension parameterized by $\lambda$ is given by the equations \eqref{gfe} and \eqref{sfe}. 

The statement of the holographic correspondence in terms of this induced theory is fairly straightforward at least at large $N$. It is just the statement that the effective action $\mathcal{S}^{ind}[g,\phi]$ satisfies the Hamilton--Jacobi equations of general relativity in one higher dimension: 
\begin{equation}
 \frac{\partial \mathcal{S}^{ind}[g,\phi]}{\partial \lambda}=-\frac{1}{\sqrt{g}}\left(G_{\mu\nu\rho\sigma}\frac{\delta \mathcal{S}^{ind}}{\delta g_{\mu\nu}}\frac{\delta \mathcal{S}^{ind}}{\delta g_{\rho\sigma}}+\frac{1}{2}\left(\frac{\delta \mathcal{S}^{ind}}{\delta \phi}\right)^{2}\right)-\sqrt{g}\left(-\frac{(\partial_{\mu}\phi\partial^{\mu}\phi)}{2}+R+V(\phi)\right)=0,
\end{equation} 
and therefore should be identified with the on shell action for the bulk theory. This was noticed first by Liu and Tseytlin in \cite{tseytliu}, where checks at the level of the linearized theory in the bulk were performed. The only novel insight here has to do with the radial development of the induced theory leading to gradient flow conditions for the theory's fundamental fields.  

\section{Conclusions and Discussion}
Thus we find that the flow triggered by the addition of certain double trace deformations to the action of four dimensional Holographic CFTs can be mapped to the Hamiltonian constraints for gravity in five dimensional AdS space through the identifications:
\begin{equation}
\sqrt{g}\langle T^{\mu\nu}\rangle-\frac{\delta S[g,\phi]}{\delta g_{\mu\nu}} =\pi^{\mu\nu}, \ \ \sqrt{g}\langle \mathcal{O}\rangle-\frac{\delta S[g,\phi]}{\delta \phi} =p_{\phi},
\end{equation}
where the functional $S[g,\phi]$ is given by \eqref{pot}, and functions in its definition satisfy the relations \eqref{suppot1}-\eqref{suppot3} and \eqref{anomfun1}-\eqref{anomfun3}. It also drives gradient flows:
\begin{equation}
\frac{1}{\mu}\frac{\partial g_{\mu\nu}}{\partial \lambda}=G_{\mu\nu\rho\sigma}\frac{\delta S[g,\phi]}{\delta g_{\rho\sigma}}, \ \ \frac{1}{\mu}\frac{\partial \phi}{\partial \lambda}=\frac{\delta S[g,\phi]}{\delta \phi},
\end{equation}
which can be seen as the equations resulting from applying the generalized gradient flow regularization procedure to the induced gravity theory obtained by integrating out the fundamental CFT fields. 

The starting point for this analysis, namely the equation
\begin{equation}
\frac{\partial \textrm{ln}Z[g,\phi]}{\partial \lambda}= \int \textrm{d}^{4}x \left(-\mu G_{\mu\nu\rho\sigma}\langle T^{\mu\nu} \rangle \langle T^{\rho\sigma} \rangle-\frac{\kappa}{2}\langle \mathcal{O}\rangle^{2}\right)+ \mathcal{A}[g,\phi],
\end{equation}
is a statement that the large $N$ holographic CFT is deformed by the stress tensor and scalar double trace deformations fails to be scale invariance due to both the expectation values of these operators in addition to the conformal anomaly, as mentioned above. However, the other assumption being made to justify this expression is that the only scales that enter the problem are those associated to these double trace deformations and no other. There is no guarantee that this should be the case, especially since the stress tensor double trace deformation, and depending on the scaling dimension of the scalar operator, also its double trace deformation are irrelevant and typically, irrelevant deformations once turned on trigger an infinite number of other increasingly irrelevant deformations. In other words, the statement that only a finite number of scales, associated to each of the possibly irrelevant deformations added to the Holographic CFT is a nontrivial assumption.

In order to successfully transform this equation into the scalar constraint for a one higher dimensional gravitational theory, we also saw how it was necessary to equate $\kappa$ and $\mu$. This can be seen as a translation of the statement of universal gravitational coupling. This means that the equivalence principle in the bulk requires a stronger assumption than the one just mentioned in the previous paragraph, namely that somehow there is really just one scale associated to the finite set of double trace deformations added to the holographic CFT. 

The fact that general covariance and the equivalence principle imply nontrivial restrictions on the properties of the flow of the dual theory on the boundary should come as no surprise. In the lower dimensional example of the $T\bar{T}$ deformed large $c$ holographic CFTs, the statement of emergent general covariance in the bulk, when translated into a property of the local renormalization group flow of the holographic theory, results a in a condition which `protects' the fact that there is just one scale associated to the irrelevant $T\bar{T}$ deformation that is introduced. See \cite{me1} for details. So covariance in the bulk seems to have some important relationship to the regularization of the holographic CFT being such that the only the scale associated to the double trace deformations are introduced. This consequently implies that the addition of a special subset of irrelevant operators doesn't trigger an infinite number of other more irrelevant operators thereby leading to a proliferation of scales. More generally, the codification of emergent of bulk diffeomorphism invariance in local Holographic RG flows is through what is called the holographic Wess--Zumino consistency condition \cite{me3}.

It would help to generalize the above construction to holographic theories in any number of dimensions, but given the necessity of the conformal anomaly and its cancellation against the `square' of the functional derivatives of counterterms in this article, a more straightforward generalization would just be to understand the role of such double trace deformations in all even dimensional holographic theories. Furthermore, the works \cite{lkrmar}, \cite{mtay} and \cite{hart} considered also vector sources which couple to gauge fields. This generalization too would be interesting to work out. 

The case of odd dimensional holographic theories is likely to be more involved because of the absence of the conformal anomaly. The linchpin to obtaining the right structure of the radial Hamiltonian in the bulk in this article was the partial cancellation between the `squares' of functional derivatives of $S[g,\phi]$ with respect to the sources $(g_{\mu\nu}, \phi) $, and the conformal anomaly in order to obtain the potential terms in the radial constraint. Without the anomaly to cancel against the higher derivative terms arising from squaring these functional derivatives, the procedure described here cannot be used to obtain the right bulk Hamitlonian.

In the two dimensional $T\bar{T}$ deformed holographic theories, the entanglement and Renyi entropies were computed at large $c$ in \cite{WV}. The entanglement entropy was found to match the length of a geodesic running between appropriately identified points on the boundary of $AdS_{3}$, and thus it seems like the Ryu--Takayanagi prescription generalizes to case of finite cutoff holography in that setting. A perturbtive calculation of the entanglement entropy for a different state and geometry was carried out in \cite{otherent}. It would also be fruitful to study the entanglement properties of the four dimensional holgoraphic CFTs deformed by the double trace operators studied in this article and check whether it is indeed the case that the Ryu--Takayanagi prescription remains valid when there is a boundary at finite radius more generally. 
\section{Acknowledgements}
I thank Rob Myers for providing detailed feedback on a previous version of this manuscript. I thank William Donnelly for useful discussions and comments on this work. I would also like to thank Lee Smolin for support and encouragement. 

This research was supported in part by the Perimeter Institute for Theoretical Physics. 
Research at Perimeter Institute is supported by the Government of Canada through the Department of Innovation, Science and Economic Development Canada and by the Province of Ontario through the Ministry of Research, Innovation and Science.
\bibliographystyle{utphys}
\bibliography{GradFlowAdS5}

\providecommand{\href}[2]{#2}\begingroup\raggedright\begin{thebibliography}{10}

\bibitem{hemPol}
I.~Heemskerk and J.~Polchinski, ``{Holographic and Wilsonian Renormalization
  Groups},'' {\em JHEP} {\bf 06} (2011) 031,
  \href{http://xxx.lanl.gov/abs/1010.1264}{{\tt 1010.1264}}.

\bibitem{flrang}
T.~Faulkner, H.~Liu, and M.~Rangamani, ``{Integrating out geometry: Holographic
  Wilsonian RG and the membrane paradigm},'' {\em JHEP} {\bf 08} (2011) 051,
  \href{http://xxx.lanl.gov/abs/1010.4036}{{\tt 1010.4036}}.

\bibitem{lecholrg}
I.~Papadimitriou, ``{Lectures on Holographic Renormalization},'' {\em Springer
  Proc. Phys.} {\bf 176} (2016) 131--181.

\bibitem{SSLee}
S.-S. Lee, ``{Quantum Renormalization Group and Holography},'' {\em JHEP} {\bf
  01} (2014) 076, \href{http://xxx.lanl.gov/abs/1305.3908}{{\tt 1305.3908}}.

\bibitem{ADM}
R.~L. Arnowitt, S.~Deser, and C.~W. Misner, ``{The Dynamics of general
  relativity},'' {\em Gen. Rel. Grav.} {\bf 40} (2008) 1997--2027,
  \href{http://xxx.lanl.gov/abs/gr-qc/0405109}{{\tt gr-qc/0405109}}.

\bibitem{mmv}
L.~McGough, M.~Mezei, and H.~Verlinde, ``{Moving the CFT into the bulk with $
  T\overline{T} $},'' {\em JHEP} {\bf 04} (2018) 010,
  \href{http://xxx.lanl.gov/abs/1611.03470}{{\tt 1611.03470}}.

\bibitem{lkrmar}
P.~Kraus, J.~Liu, and D.~Marolf, ``{Cutoff AdS$_{3}$ versus the $ T\overline{T}
  $ deformation},'' {\em JHEP} {\bf 07} (2018) 027,
  \href{http://xxx.lanl.gov/abs/1801.02714}{{\tt 1801.02714}}.

\bibitem{gir}
G.~Giribet, ``{$T\bar{T}$-deformations, AdS/CFT and correlation functions},''
  {\em JHEP} {\bf 02} (2018) 114,
  \href{http://xxx.lanl.gov/abs/1711.02716}{{\tt 1711.02716}}.

\bibitem{giv}
M.~Asrat, A.~Giveon, N.~Itzhaki, and D.~Kutasov, ``{Holography Beyond AdS},''
  {\em Nucl. Phys.} {\bf B932} (2018) 241--253,
  \href{http://xxx.lanl.gov/abs/1711.02690}{{\tt 1711.02690}}.

\bibitem{Nima}
G.~Bonelli, N.~Doroud, and M.~Zhu, ``{$T \bar{T}$-deformations in closed
  form},'' {\em JHEP} {\bf 06} (2018) 149,
  \href{http://xxx.lanl.gov/abs/1804.10967}{{\tt 1804.10967}}.

\bibitem{mtay}
M.~Taylor, ``{TT deformations in general dimensions},'' {\em
  arXiv:1805.10287[hep-th]} (2018)
  \href{http://xxx.lanl.gov/abs/1805.10287}{{\tt 1805.10287}}.

\bibitem{hart}
T.~Hartman, J.~Kruthoff, E.~Shaghoulian, and A.~Tajdini, ``{Holography at
  finite cutoff with a $T^2$ deformation},'' {\em arXiv:1807.11401[hep-th]}
  (2018) \href{http://xxx.lanl.gov/abs/1807.11401}{{\tt 1807.11401}}.

\bibitem{fzam}
F.~A. Smirnov and A.~B. Zamolodchikov, ``{On space of integrable quantum field
  theories},'' {\em Nucl. Phys.} {\bf B915} (2017) 363--383,
  \href{http://xxx.lanl.gov/abs/1608.05499}{{\tt 1608.05499}}.

\bibitem{tat}
A.~Cavaglià, S.~Negro, I.~M. Szécsényi, and R.~Tateo, ``{$T
  \bar{T}$-deformed 2D Quantum Field Theories},'' {\em JHEP} {\bf 10} (2016)
  112, \href{http://xxx.lanl.gov/abs/1608.05534}{{\tt 1608.05534}}.

\bibitem{bunster}
C.~Teitelboim, ``{How commutators of constraints reflect the space-time
  structure},'' {\em Annals Phys.} {\bf 79} (1973) 542--557.

\bibitem{Skenderis}
M.~Henningson and K.~Skenderis, ``{The Holographic Weyl anomaly},'' {\em JHEP}
  {\bf 07} (1998) 023, \href{http://xxx.lanl.gov/abs/hep-th/9806087}{{\tt
  hep-th/9806087}}.

\bibitem{BK}
V.~Balasubramanian and P.~Kraus, ``{A Stress tensor for Anti-de Sitter
  gravity},'' {\em Commun. Math. Phys.} {\bf 208} (1999) 413--428,
  \href{http://xxx.lanl.gov/abs/hep-th/9902121}{{\tt hep-th/9902121}}.

\bibitem{dilax}
I.~Papadimitriou, ``{Holographic Renormalization of general dilaton-axion
  gravity},'' {\em JHEP} {\bf 08} (2011) 119,
  \href{http://xxx.lanl.gov/abs/1106.4826}{{\tt 1106.4826}}.

\bibitem{counterright}
M.~Bianchi, D.~Z. Freedman, and K.~Skenderis, ``{Holographic
  renormalization},'' {\em Nucl. Phys.} {\bf B631} (2002) 159--194,
  \href{http://xxx.lanl.gov/abs/hep-th/0112119}{{\tt hep-th/0112119}}.

\bibitem{grg}
S.~Jackson, R.~Pourhasan, and H.~Verlinde, ``{Geometric RG Flow},'' {\em
  arXiv:1312.6914[hep-th]} (2013) \href{http://xxx.lanl.gov/abs/1312.6914}{{\tt
  1312.6914}}.

\bibitem{Nakayama1}
Y.~Nakayama, ``{$a - c$ test of holography versus quantum renormalization
  group},'' {\em Mod. Phys. Lett.} {\bf A29} (2014), no.~29 1450158,
  \href{http://xxx.lanl.gov/abs/1401.5257}{{\tt 1401.5257}}.

\bibitem{Osborn1}
H.~Osborn, ``{Weyl consistency conditions and a local renormalization group
  equation for general renormalizable field theories},'' {\em Nucl. Phys.} {\bf
  B363} (1991) 486--526.

\bibitem{twoder}
E.~Kiritsis, W.~Li, and F.~Nitti, ``{Holographic RG flow and the Quantum
  Effective Action},'' {\em Fortsch. Phys.} {\bf 62} (2014) 389--454,
  \href{http://xxx.lanl.gov/abs/1401.0888}{{\tt 1401.0888}}.

\bibitem{farkas}
S.~Farkas and E.~J. Martinec, ``{Gravity from the Extension of Spatial
  Diffeomorphisms},'' {\em J. Math. Phys.} {\bf 52} (2011) 062501,
  \href{http://xxx.lanl.gov/abs/1002.4449}{{\tt 1002.4449}}.

\bibitem{me2}
H.~Gomes and V.~Shyam, ``{Extending the rigidity of general relativity},'' {\em
  J. Math. Phys.} {\bf 57} (2016), no.~11 112503,
  \href{http://xxx.lanl.gov/abs/1608.08236}{{\tt 1608.08236}}.

\bibitem{tseytliu}
H.~Liu and A.~A. Tseytlin, ``{D = 4 superYang-Mills, D = 5 gauged supergravity,
  and D = 4 conformal supergravity},'' {\em Nucl. Phys.} {\bf B533} (1998)
  88--108, \href{http://xxx.lanl.gov/abs/hep-th/9804083}{{\tt hep-th/9804083}}.

\bibitem{wilfow}
M.~Lüscher, ``{Properties and uses of the Wilson flow in lattice QCD},'' {\em
  JHEP} {\bf 08} (2010) 071, \href{http://xxx.lanl.gov/abs/1006.4518}{{\tt
  1006.4518}}. [Erratum: JHEP03,092(2014)].

\bibitem{aokisan0}
S.~Aoki, K.~Kikuchi, and T.~Onogi, ``{Generalized Gradient Flow Equation and
  Its Applications},'' {\em PoS} {\bf LATTICE2015} (2016) 305,
  \href{http://xxx.lanl.gov/abs/1511.06561}{{\tt 1511.06561}}.

\bibitem{floweq1}
S.~Aoki, J.~Balog, T.~Onogi, and P.~Weisz, ``{Flow equation for the large $N$
  scalar model and induced geometries},'' {\em PTEP} {\bf 2016} (2016), no.~8
  083B04, \href{http://xxx.lanl.gov/abs/1605.02413}{{\tt 1605.02413}}.

\bibitem{floweq2}
S.~Aoki, J.~Balog, T.~Onogi, and P.~Weisz, ``{Flow equation for the scalar
  model in the large $N$ expansion and its applications},'' {\em PTEP} {\bf
  2017} (2017), no.~4 043B01, \href{http://xxx.lanl.gov/abs/1701.00046}{{\tt
  1701.00046}}.

\bibitem{aokisan1}
S.~Aoki and S.~Yokoyama, ``{AdS geometry from CFT on a general conformally flat
  manifold},'' {\em Nucl. Phys.} {\bf B933} (2018) 262--274,
  \href{http://xxx.lanl.gov/abs/1709.07281}{{\tt 1709.07281}}.

\bibitem{aokisan2}
S.~Aoki and S.~Yokoyama, ``{Flow equation, conformal symmetry, and anti-de
  Sitter geometry},'' {\em PTEP} {\bf 2018} (2018), no.~3 031B01,
  \href{http://xxx.lanl.gov/abs/1707.03982}{{\tt 1707.03982}}.

\bibitem{aokisan3}
S.~Aoki, J.~Balog, and S.~Yokoyama, ``{Holographic computation of quantum
  corrections to the bulk cosmological constant},'' {\em
  arXiv:1804.04636[hep-th]} (2018)
  \href{http://xxx.lanl.gov/abs/1804.04636}{{\tt 1804.04636}}.

\bibitem{me1}
V.~Shyam, ``{Background independent holographic dual to $T\bar{T}$ deformed CFT
  with large central charge in 2 dimensions},'' {\em JHEP} {\bf 10} (2017) 108,
  \href{http://xxx.lanl.gov/abs/1707.08118}{{\tt 1707.08118}}.

\bibitem{me3}
V.~Shyam, ``{Connecting holographic Wess-Zumino consistency condition to the
  holographic anomaly},'' {\em JHEP} {\bf 03} (2018) 171,
  \href{http://xxx.lanl.gov/abs/1712.07955}{{\tt 1712.07955}}.

\bibitem{WV}
W.~Donnelly and V.~Shyam, ``{Entanglement entropy and $T \overline{T}$
  deformation},'' {\em Phys. Rev. Lett.} {\bf 121} (2018) 131602,
  \href{http://xxx.lanl.gov/abs/1806.07444}{{\tt 1806.07444}}.

\bibitem{otherent}
B.~Chen, L.~Chen, and P.-X. Hao, ``{Entanglement entropy in
  $T\overline{T}$-deformed CFT},'' {\em Phys. Rev.} {\bf D98} (2018), no.~8
  086025, \href{http://xxx.lanl.gov/abs/1807.08293}{{\tt 1807.08293}}.

\end{thebibliography}\endgroup

\begin{comment}

\end{comment}

\end{document}